\documentstyle[12pt]{article}
\begin{document}

\newcommand{\ep}{\epsilon}
\newcommand{\be}{\begin{eqnarray}}
\newcommand{\ee}{\end{eqnarray}}
\newcommand{\bea}{\begin{eqnarray*}}
\newcommand{\eea}{\end{eqnarray*}}
\newcommand{\bh}{\beta \hbar}
\newcommand{\uv}{\mu\nu}
\newcommand{\mn}{\mu \nu}
\newcommand{\m}{\mu}
\newcommand{\n}{\nu}
\newcommand{\h}{\hbar}

\rightline{{\bf CWRU-P17-96}}
\rightline{November 1996}
\baselineskip=16pt
\vskip 0.5in
\begin{center}
{\bf\large
QUANTUM HAIR, INSTANTONS, AND BLACK HOLE THERMODYNAMICS: SOME NEW RESULTS}
\end{center}
\vskip0.2in
\begin{center}
Lawrence M. Krauss \footnote{also Department of Astronomy} and Hong Liu
\vskip .1in
{\small\it Department of Physics \\
 Case Western Reserve University\\
10900 Euclid Ave.,
Cleveland, OH 44106-7079}

\vskip 0.4in
\end{center}

\begin{abstract}

We present results obtained by a consideration
 of the non-classical energy
momentum tensor associated with Euclidean Instantons outside the event 
horizon of black holes. We demonstrate here how this allows an
analytic estimate to be made of the effect of discrete quantum hair on the
temperature of the black hole, in which the role of violations of the weak
energy condition associated with instantons is made explicit, and in which the
previous results of Coleman, Preskill, and Wilczek are extended.  Last, we
demonstrate how the existence of a non-classical electric field outside the
event horizon of black holes, uncovered by these authors, can be identified
with a well-known effect in the Abelian-Higgs model in two dimensions.  In this
case, there is a one-to-one connection between the discrete charge of a black
hole and a topological phase in two dimensions.  

\end{abstract}

\newpage
\baselineskip=21pt

\section{Introduction}

Our
understanding of how gravity and quantum mechanics may be reconciled has
changed considerably as a result of semiclassical considerations.
Arguably the most famous example involves Hawking radiation, 
with its associated
implications for black hole thermodynamics. Indeed, the very process of black
hole evaporation, which is made possible in principle by the effects of quantum
fluctuations near the event horizon of a black hole, has challenged the
foundations of quantum mechanics itself.  If Hawking radiation by a black hole
is purely thermal, with no hidden correlations between outgoing emitted
particles, then complete evaporation of a black hole would allow pure quantum
mechanical states to evolve into mixed states, violating unitarity, and leading
to the loss of information (i.e.\cite{hawking,preskill}) 

Much work has been carried out in the past decade aimed at
addressing this potential problem. No definitive solution is yet at hand, but
a lot has been learned during the process about such things as state counting
and black hole entropy, possible new Planck Scale phenomena, and phenomena
in higher dimensions.  

At the same time, several interesting results have been obtained which involve
essentially only phenomena far below the Planck Scale.  In particular, it
has recently been established that
black holes can harbor ``quantum hair" (i.e.\cite{bowick,krauss,CPWa})---that 
is,
quantum mechanical observables can be associated with black holes beyond
those allowed by the classical ``no hair theorems" \cite{bek,teit,adler}
(for recent discussions, see \cite{bek2}).

We first briefly describe a canonical example of gauged quantum hair
\cite{krauss}.  Consider an abelian U(1) gauge theory containing
two matter fields
$\eta$ and $\phi$ with charge $Ne$ and $e$ respectively.  If the field $\eta$
condenses at some energy scale $v$, then the gauge field will become
massive by the Higgs mechanism, and below this scale the effective theory
involving only the light field $\phi$ will have a residual discrete $Z_N$
symmetry.

At the same time, this 
low energy broken symmetry theory also can contain stable strings
threaded by magnetic flux $2\pi /Ne$.  The scattering of $\phi$ quanta, with
charge $e$ from such strings is dominated the Aharonov-Bohm effect
\cite{bohm,wilz}.  Since such scattering 
involves quantum phases uniquely determined by the product of
charge and flux, a determination of the total charge
modulo N which scatters off the string is thus possible.
Since the quantum phases in question are global quantities, a $\phi$ quanta
which falls into a black hole will be measurable as such even after it falls
inside the event horizon of the black hole \cite{krauss,presk,wilz2,ckpw}.  
Such
a charged black hole therefore has ``quantum hair".

The proof of the existence of non-classical gauge hair on black holes
\cite{krauss,presk,CPWa} which could obviate the famous no-hair theorems
for black holes led to hopes that such non-classical hair might alleviate the
black hole information loss problem. However, while such hopes evaporated about
as quickly as a small black hole might, efforts soon focussed on exploring
whether quantum hair might have any other observable effects.

Indeed, given the semiclassical relationship between entropy, area, and
temperature for black holes, it is natural to expect that any
restriction on the number of microstates associated with a
given classical black hole macrostate, such as would occur if one could measure
additional black hole quantum numbers, would have a related effect on the
black hole's entropy and temperature.  This possibility
was explored most recently in a beautiful series of
papers by Coleman, Preskill and Wilczek
\cite{CPWa,CPWb}.  These authors 
uncovered an (exponentially small) effect on
temperature, but surprisingly the sign of the temperature change depended upon
the relative scale of the spontaneous symmetry breaking associated with the
quantum hair compared to 
the inverse size of the black hole event horizon.  More interestingly perhaps,
they also uncovered a new observable associated with quantum hair: a
non-classical electric field could exist and be measured outside of the
event horizon.

Since these calculations were performed in the context of the 
Euclidean Action in
curved space, it would be of some interest to examine these effects in a
formalism more closely related to Minkowski space methods, 
where presumably some additional physical insights might arise.   Recently,
we outlined such an approach \cite{KLHprl}.
Here we extend
the arguments outlined in our earlier letter, and also provide a more detailed
derivation of both our original and several new results. In
section 1 we outline our formalism.  In section 2 we review the
application of this method to the case of a Reissner-Nordstrom black hole.  In
the next section we discuss the Abelian Higgs theory in the thin string
limit, and then follow this with a detailed discussion of the thick string
limit, where several additional subtleties arise.   Finally we describe how
the results obtained by CPW can be recast in the light of similar results from
two dimensional quantum field theory.  In this case, there is a one to one
mapping between discrete charge on a black hole, and a topological phase.  We
conclude with a brief discussion of possible implications of our results.
 
\section{Formalism}

Following CPW's demonstration that a non-classical field exists outside the
event horizon of a black hole endowed with discrete gauge hair, 
it then seems reasonable 
to focus on the energy momentum tensor outside the black hole.  
 A Minkowski-space formalism which does just this was developed by Visser
\cite{viss} for treating "dirty" black holes, where non-zero 
matter fields exist
outside the event horizon.  Unfortunately, 
since the electric field associated with discrete hair is 
non-classical---i.e.it is not a solution of the coupled vacuum Einstein-Maxwell
equations--- standard methods such as Visser's, which require such solutions, 
cannot be applied
directly.

It is true, however, that while
the electric field generated outside the event horizon is not
a solution of the Minkowski field equations, the individual instantons whose
contributions sum to produce such a field are solutions of the coupled
Euclidean Einstein-Maxwell equations.  As a result, we have
utilized \cite{KLHprl} a hybrid approach, in
which we reproduce the Visser formalism in Euclidean space, and then
focus on the effect of individual instantons.  We derive here several key
results which we will then employ to analyze a variety of black hole
configurations with and without discrete hair.

The Euclidean spacetime metric
generated by a static spherically symmetric  distribution of matter can be put
in the form:
\[
ds^{2} = e^{-2\phi(r)} ( 1 - \frac{b(r)}{r}) dt^{2} 
         + ( 1 - \frac{b(r)}{r})^{-1} dr^{2} + r^{2}d \Omega^{2}
\]
With the assumption  that the metric has an asyptotically 
flat geometry and an event horizon, boundary conditions 
can be imposed as:
\[
\phi(\infty) = 0, \,\,\,\, b(\infty) = 2GM_{BH}, \,\,\,\, b(r_{H}) = r_{H}
\]
where $M_{BH}$  is the mass of the  black hole and $r_{H}$ is the horizon
size.
Einstein's equations,
\[
R_{\mu\nu} - \frac{1}{2}g_{\mu\nu}R = 8 \pi GT_{\mu\nu}
\]
with
\[
T_{\mu\nu} = \frac{2}{\sqrt{g}}
                  \frac{\partial ({\cal L}_{E} \sqrt{g})}{\partial g^{\mu\nu}}
\]
can then be solved formally to give $b(r)$ and $\phi(r)$ in 
terms of the components of the energy momentum tensor:
\be
b(r) = r_{H} - 8 \pi G \int_{r_{H}}^{r}dr' \rho(r') r'^{2} 
\label{eq:b} \\
\phi(r) =  4\pi G \int_{r}^{\infty}dr'  
       \,\frac{(\tau - \rho)r'}{1- b(r')/r'} 
\label{eq:phi} 
\ee
In the above we define
\[
T_{t}^{t} = \rho, \,\,\,\, T_{r}^{r} = \tau, \,\,\,\,
T_{\theta}^{\theta} = T_{\varphi}^{\varphi} = - \mu.
\]
which satisfy the conservation law  
\be
\tau' + (\tau - \rho)( -\phi' + \frac{1}{2} 
(\ln(1- \frac{b(r)}{r}))' ) + \frac{2}{r} (\mu + \tau) = 0
\label{eq:cons}
\ee
Using  equations (\ref{eq:b}) (\ref{eq:phi}), the Hawking 
temperature and 
the horizon size of the black hole can now be expressed as 
(i.e. see \cite{viss}):
\begin{eqnarray}
\frac{1}{\beta\hbar} & = & \frac{1}{4 \pi r_{H}}e^{- \phi (r_{H})} 
                     (1 - b'(r_{H})) \\
 & = & \frac{1}{4 \pi r_{H}} \exp( -4\pi G \int_{r_{H}}^{\infty}dr  
       \,\frac{(\tau - \rho)r}{1- b(r)/r} ) 
        (1 + 8 \pi G \rho_{H}r_{H}^{2})  
\label{eq:HT} \\
r_{H} & = & 2 G M_{BH} + 8 \pi G \int_{r_{H}}^{\infty}dr \rho r^{2}
\label{eq:rh}
\end{eqnarray}

When the external matter contribution to the geometry is much 
smaller 
than that of the black hole,
equations  (\ref{eq:HT}) and (\ref{eq:rh})
can be systematically expanded and the  lowest order
corrections to
 black hole thermodynamics 
can then be obtained.   
Now define 
\begin{eqnarray*}
A & = & \frac{8 \pi G}{2} \int_{r_{H}}^{\infty} \frac{\rho - \tau}  
{r-r_{H}} r^{2} dr \\
B & = & 8 \pi G \rho_{H} r_{H}^{2} \\
m & = & 4 \pi \int_{r_{H}}^{\infty} (2\mu + \rho -\tau) r^{2} dr
\end{eqnarray*}
Expanding (\ref{eq:HT}) and (\ref{eq:rh}) to first
order in 
$A$, $B$, and $(m/M_{BH})$ and using the conservation law 
(\ref{eq:cons})
we reach:
\be
\frac{1}{\beta\hbar} & = & \frac{1}{4\pi}\frac{2GM_{BH}}{r_{H}^{2}}
(1 + \frac{m}{M_{BH}}) + higher \,\, order \,\, corrections 
\label{eq:HT1}\\
r_{H} & = & 2GM_{BH} \frac{1 + \frac{m}{M_{BH}}}{1+A+B}
+ higher \,\, order \,\, corrections 
\label{eq:rh1}
\ee
(Note that (\ref{eq:rh1}) is equivalent to the Bardeen-Carter-Hawking mass 
theorem \cite{bardeen}.)

Plugging  equation (\ref{eq:rh1}) into (\ref{eq:HT1}), we get  
an expression for $\beta\hbar$ 
in terms of  only the components of the energy momentum
tensor and $M_{BH}$ \cite{KLHprl}: ( In this order,  $r_{H}$ can be replaced 
by its lowest
order piece, i.e.  $r_{H} = 2GM_{BH}$.)
\be
\beta\hbar = 8\pi GM_{BH} ( 1 + \frac{m}{M_{BH}} - 2(A+B) + 
higher \,\, order \,\, corrections).
\label{eq:HTf}
\ee
For further reference, it is useful to write out this expression in more 
detail.
In particular, 
\be
\frac{m}{M_{BH}} -2(A+B) = -\frac{ 8\pi G}{r_{H}} \int_{r_{H}}^{\infty} dr
[ (4\mu r r_{H} - 2\mu r^{2}) - (\rho - \tau)r(r-r_{H})]
\label{eq:app}
\ee

It is clear that the sign of the
correction to the
black hole temperature, for fixed mass, depends upon the relative sign of the
term 
$m/M_{BH} -2 (A+B)$.  Using this fact, we can
further explore the result of CPW that 
discrete charge on a black hole can either raise or lower the black hole
temperature, depending upon the ratio of the gauge symmetry breaking scale
to the inverse horizon size of the black hole, a somewhat surprising result at
first sight. Indeed, this is perhaps even more surprising when
one considers that all forms of classical matter which satisfy the
Weak Energy Condition (WEC)
also satisfy the relation $m/M_{BH} -2 (A+B) \ge 0$.  This is one
example of a general result that
classical matter outside a black hole can only lower its 
temperature\cite{viss}. By framing the discussion in terms
of the quantity $m/M_{BH} -2 (A+B) $, our analysis will allow us
to establish an explicit connection between the WEC and
the effects of instantons, about which we will have more to say later. 

 In the following sections we
use the above results to examine the effect of various instanton contributions
to the temperature of a black hole in the semiclassical limit.  Following
Coleman, Preskill, and Wilczek (CPW), we explore first the case of an unbroken
U(1) theory with unshielded electric charge, 
and then the broken theory involving
discrete hair.   A great deal of physical insight can be gained by focussing
on the sometimes subtle distinctions between these cases.  

\section{The Euclidean Reissner-Nordstrom Black Hole}

A Euclidean Reissner-Nordstrom black hole provides a simple
application of the above formalism, and one to which we shall
be able to compare our later results. ( In what follows, we 
will assume $M_{pl}/M_{BH} \ll 1$.)  We recall that this is 
a Euclidean solution of the field equations which is periodic in
imaginary time with period $\beta\hbar$.  Moreover, because it describes
a configuration with finite charge, the temporal component of the
gauge field is restricted by a gauge constraint ( i.e. see \cite{CPWa}):

\bea
e \int_{0}^{\bh} dt \,\, A_{t}|_{r=\infty} = \omega
\eea
where $\omega$ is a constant related to the value of the electric
charge, as we will explicitly derive momentarily.

The
action of this system is:
\[
S_{E} = \int d^{4}x \sqrt{g}\, ({\cal L}_{em} - {1\over 16\pi G}R)
 + {1 \over 2} (\beta\hbar) M_{BH}
\]
with
\[
{\cal L}_{em} = {1 \over 16 \pi} \, 
g^{\mu \lambda} g^{\nu \sigma} F_{\mu\nu}F_{\lambda\sigma} 
\]
The stationary point solution is:
\bea
A_{t} = & {\omega \over \bh e}(1- {r_{H} \over r}) \\
F_{rt} = & {\omega \over \bh e} {r_{H} \over r^{2}} \\
\phi(r) = & 0 
\eea
The energy momentum tensor is given by: 
\be
\rho = \tau = \mu = { 1 \over 8 \pi} F_{rt}^{2} 
\propto { 1 \over r^{4}} \label{eq:em}
\ee

Now, interestingly, from (\ref{eq:app}),
\be
2(A+B) - \frac{m}{M_{BH}}  =  \frac{ 8\pi G}{r_{H}} 
\int_{r_{H}}^{\infty} dr (4\mu r r_{H} - 2\mu r^{2}) = 0
\label{eq:emzero}
\ee
%
Thus, in this case in order to derive the correction to the
black hole temperature resulting from the classical electric field outside the
horizon, we need go to higher orders in
$A$,
$B$ and
$m/M_{BH}$. 

Of course, in the case of unbroken electromagnetism one
can derive an exact result. Plugging equation (\ref{eq:em}) into equations
(\ref{eq:HT}) and (\ref{eq:rh}), we get
\bea 
\bh = 4 \pi r_{H} { 1 \over 1 + G \Phi} \\
r_{H} = 2 G M_{BH} { 1 \over 1 - G \Phi}
\eea
with  $ \Phi$ defined by
\[
\Phi =   ({\omega \over \bh e})^{2} 
\]
In this case, $\Phi$ is the electric potential at the horizon, and
the above equation then gives the relation between $\omega$ and the
electric charge $Q$.

From this, it is straightforward to derive: 
\bea
M_{BH} = {\bh \over 8 \pi G} ( 1- G^{2} \Phi^{2})
\eea
and thus
\bea
\bh & \approx & 8 \pi G M_{BH} 
( 1  + ({\omega \over 8 \pi e})^{4} {1 \over G^{2}M_{BH}^{4}} 
+ \cdots ) \\
& \sim & 8 \pi G M_{BH} 
( 1 + O({M_{pl}^{4} \over M_{BH}^{4}}) + \cdots )
\eea

This is the Euclidean version of the
well known result that classical electrically charged black holes have
a lower temperature, for fixed mass, than their uncharged counterparts.

Note that the action can
also 
be evaluated to lowest order in $G\Phi$, or equivalently $(M_{pl}/M_{BH})^2$,
(using the fact that $R=0$ outside the event horizon):
\bea
S & = & {(\bh)^{2} \over 16 \pi G} 
[ 1 + G ({\omega \over \bh e})^{2} ]^{2} \\
& \sim & {(\bh)^{2} \over 16 \pi G} + O(1) 
+ O[{M_{pl}^{2} \over M_{BH}^{2}}] + \cdots 
\eea
This result will also have some significance later in our analysis.

\section{Discrete Charge and the Effect of Quantum Hair}

Now we consider the Euclidean Einstein-Abelian-Higgs system which, as
described earlier, provides the prototypical example of quantum hair.

The action of this system is: 
\[
S_{E} = \int d^{4}x \sqrt{g}\, ({\cal L}_{ah} - {1\over 16\pi G}R)
 + {1 \over 2} (\beta\hbar) M_{BH}
\]
with 
\[
{\cal L}_{ah} = {1 \over 4 \pi} 
[\, \frac{1}{4}g^{\mu \lambda}g^{\nu \sigma}
F_{\mu\nu}F_{\lambda\sigma} 
  + g^{\uv}(D_{\mu}\phi)^{*}(D_{\nu}\phi) 
  + {\lambda \over 4}(|\phi|^{2}-v^{2})^{2} \,]
\]

The above action has solutions corresponding to a vortex sitting 
in the 2-d Euclidean $r-t$ plane of a black hole 
(i.e. see also \cite{gregory}).
The two other Euclidean dimensions $\theta , \phi$, (which would correspond to 
$z,t$ for a corresponding vortex in Minkowski space) are suppressed.  
As emphasized by CPW, in
a Euclidean path integral formalism these
instanton solutions play a central role in
producing the observable effects of discrete charge outside
of the black hole event horizon, as the sum over these instantons includes
Aharonov-Bohm phases which are sensitive to the discrete charge contained 
the black hole. 
As we have described earlier, we can explore the effects on temperature
of the individual
instantons utilizing our formalism.
  
We use standard ansatz for these vortices:
\bea
\phi = v f(r) e^{-i{2 \pi \over \bh}t}, \,\,\,\, \\
A_{t} = {2 \pi \over \bh}{1 \over e}( 1 - a(r)), \,\,\,\,
\eea
with boundary conditions:
\bea
f(r_H) = 0, \,\,\,\,
f( \infty ) = 1 \\
a(r_H) = 1, \,\,\,\,
a(\infty) = 0 
\eea
Now $A_{t}$ satifies an equation similar to that presented for
the Reissner-Nordstrom solution.  However, in this case, the 
quantization condition comes about simply as a reflection of the flux
quantization condition for vortices in the broken phase:
\bea
e \int_{0}^{\bh} dt \,\, A_{t}|_{r=\infty} = 2 \pi
\eea
(in other words, $\omega =2\pi$, rather than being a free parameter
as it was in the Reissner-Nordstrom case).

We can also write down the action for the Higgs sector:
\bea
S_{vortex} & = & 4 \pi (\bh) \int_{r_{H}}^{\infty }
 r^{2} dr e^{-\phi(r)}[ {1 \over 2}( {2 \pi \over \bh e})^{2}
a'^{2}(r) e^{2\phi(r)} \\
 & + &  (1-{ b(r) \over r })^{-1} e^{2\phi(r)}v^{2} 
( {2 \pi \over \bh})^{2} a^{2} f^{2} \\
& + &  (1-{ b(r) \over r }) v^{2} f'^{2}(r) +
{ \lambda \over 4} v^{4} (f^{2} -1 )^{2} ]
\eea

Following CPW one can consider the two limiting cases of the above 
action, depending upon whether the vortex width is much larger or smaller
than the size of the event horizon.  Equations (\ref{eq:HTf}) and 
(\ref{eq:app})  lend themselves directly to such an analysis.  
Competition among the different terms as their
$r$-dependence varies, can lead, in different limits, to a different sign for
the correction to the black hole temperature. However, what actually occurs
depends subtlely yet crucially on diffferences between the vortex solution in
flat and curved space, as we shall describe in detail here.

\subsection{Thin String Limit}

In
the thin string limit, the vortex width
$r_{s}
\ll r_{H}$, so in  (\ref{eq:app}),
\bea
2(A+B) - \frac{m}{M_{BH}} & \approx & \frac{ 8\pi G}{r_{H}} 
\int_{r_{H}}^{\infty} (4\mu r r_{H} - 2\mu r^{2}) dr \\
& \approx & 16\pi G r_{H} \int_{r_{H}}^{\infty}  \mu dr \,
\eea
Thus the correction to $\bh$ due
to the vortex instanton can  be expressed as: 
$$
\bh = 8\pi GM_{BH} [ 1 -
16\pi G r_{H} \int_{r_{H}}^{\infty}  \mu dr  ]\,
$$

In this limit, the vortex lies in the 
region $r \sim r_{H}$, where the Euclidean 
Schwartzchild metric
$$ds^{2} = ( 1 - \frac{r_{H}}{r}) dt^{2} 
         + ( 1 - \frac{r_{H}}{r})^{-1} dr^{2} + r^{2}d \Omega^{2}$$
can be well approximated by (note that our use of $\rho$ and $\tau$ in this
specific example follows standard conventions, and these co-ordinate
quantities are not to be confused with the quantities $\rho$ and $\tau$ used
elsewhere throughout this article): 
\[
ds^{2} = \rho^{2} d\tau^{2} 
         +  d \rho^{2} + r_{H}^{2} d \Omega^{2}
\]
with 
\[
\rho^{2} = 4 r_{H} (r -r_{H}), \,\,\,\,\,
\tau = {2 \pi \over \bh}t
\]

In the  space with coordinates $\rho$ and $\tau$, the vortex
actually lives in flat space and 
$$
r_{H} \int_{r_{H}}^{\infty} \mu(r) dr 
\stackrel{ \rho_{2} = 4 r_{H} (r -r_{H})}{\longrightarrow} 
{ 1 \over 2}\int_{0}^{\infty} \mu(\rho) \rho d\rho
={1 \over 4\pi} T_{vortex} \sim v^{2} 
$$
where $ \mu(\rho)$ is just the energy density of the 
Nielsen-Oleson vortex and  $T_{vortex}$ is
the tension of the vortex in flat
space. 
It is then straightforward to see that:
\bea
A & \sim & B \sim  {m \over M_{BH}} \sim {1 \over  M_{BH}}
\int_{r_{H}}^{\infty } \mu  r^{2} dr \\
& \sim & 
{r_{H}^{2} \over  M_{BH}} \int_{r_{H}}^{\infty} \mu dr  \\
& \sim &
{r_{H} \over M_{BH}} \int_{0}^{\infty} \mu(\rho) \rho d\rho \\
 & \sim  & Gv^{2} \sim {v^{2} \over M_{pl}^{2}}
\eea
which is consistent with the expansion leading to (\ref{eq:HTf}).
An estimate of $r_{s}$  follows from
$$  r_{H}r_{s} \sim \rho_{s}^{2} \sim {1 \over v^{2}} \rightarrow 
r_{s} \sim {1 \over v^{2}r_{H}}$$
where $\rho_{s}$ is the width of the vortex in the $\rho$ coordinate.

So the Hawking temperature can finally be expressed as:
\bea
\bh & =& 8\pi GM_{BH} [ 1 -
16\pi G r_{H} \int_{r_{H}}^{\infty}  \mu dr ] \\
 & =&  8\pi GM_{BH} [ 1 - 4 G T_{vortex}] \\
& \sim &  8\pi GM_{BH} [ 1 - O({v^{2} \over M_{pl}^{2}})]  \\
\eea

The
effect of a single instanton in this limit is thus to raise the black
hole temperature. This is the first
explicit example of the fact that a
semiclassical contribution by itself can violate the Weak Energy
Condition, described earlier.  That this instanton contribution violates
the WEC is perhaps not so surprising.  In the thin string case, the
symmetry breaking scale is much smaller than the event horizon size so
that effectively for all distances outside the horizon the symmetry is
completely broken, and no classical remnants of the underlying U(1) symmetry
should be visible on such scales.  We shall later contrast this to the 
thick string case.

    Of course we remind the reader that the limitation of
our analysis is that one can only treat a single
instanton (a solution of the coupled Einstein-Maxwell equations) contribution 
and not a sum over instantons (not such a solution).  
The latter sum is required
in order  to find the net thermal effect of discrete charge on the black hole, 
as
CPW did, by using the path integral summation with instantons weighted by the
appropriate action.  If one does the summation a la CPW, the
interference between instantons and anti-instantons produces a net effect
on the black hole temperature which is opposite in sign to that for a 
single instanton

Nevertheless,
the single instanton contribution to the temperature 
which we calculate using the
energy momentum formalism directly is identical with that determined by CPW in
the thin string case based on estimating the deficit angle and the instanton
action. With this in mind, and for later comparison
purposes, we also  present our estimate of the action of the vortex solution in
this limit.

\bea
S & = & \int d^{4}x \sqrt{g}\, ({\cal L}_{E} - {1\over 16\pi G}R)
+ {1 \over 2}(\bh)M_{BH} \\
 & \doteq  & {1 \over 2} \int d^{4}x \sqrt{g}\, (\rho + \tau) + 
{1 \over 2}(\bh) {\bh \over 8 \pi G} 
( 1 + 16\pi G r_{H} \int_{r_{H}}^{\infty}  \mu dr) \\
 & \doteq  & 
2 \pi \bh r_{H}^{2} \int_{r_{H}}^{\infty} (\rho + \tau) dr + 
{(\bh)^{2} \over 16 \pi G} 
( 1 + 16\pi G r_{H} \int_{r_{H}}^{\infty}  \mu dr)\\
& \doteq & {(\bh)^{2} \over 16 \pi G} + 
{1 \over 2}(\bh)^{2} r_{H} \int_{r_{H}}^{\infty} 
(2 \mu + \rho + \tau) dr \\
& \sim &  {(\bh)^{2} \over 16 \pi G}
( 1 + O({v^{2} \over M_{pl}^{2}})) 
\eea
In above, we have used  $R = -8 \pi T_{\mu}^{\mu}$.
As $\mu > |\rho|, |\tau|$, and because $\mu \sim {\cal L}_{E} >0$,
 we note that 
$S > S_{Schwarzchild}$, as one would expect for the instanton 
approximation to be stable.

\subsection{Thick String Limit}

The simplication which made the consideration
of the thin string limit so straightforward was that one
could simply picture the instanton as a vortex living in the two dimensions
of
a flat $r-t$ plane.  The thick string limit, in which
$ r_s >> r_H$, is much more subtle.  In this limit the 
curvature
associated with the sphere at the event horizon cannot be ignored. 
If one were to ignore this curvature and just
continue utilizing the flat space approximation one could then use well known
properties of standard flat space vortex solutions inside the core,
where the symmetry is unbroken, along with the boundary conditions associated
with the magnetic flux carried in the core, to examine eq. (10), and estimate
the instanton contribution.  

An anomaly arises in this case, however, which signals that such a procedure is
inconsistent.  We have thus far presented the Euclidean action of each
instanton solution.  If we were to calculate the action of a flat
space thick vortex, which we will present in more detail later, 
we would obtain a
result which is nonsensical---namely that the instanton action is less than the
Schwarzchild action. 
There is another worrisome result. The lowest order
correction one would find to the black hole temperature in this case is
proportional to
$v^2/M_{pl}^2$, as it was in the thin string case, while CPW argue that a 
zeroth
order contribution in this limit dominates.  

Remarkably, as we now demonstrate explicitly, a proper consideration of the
full curved space vortex solution in the thick string limit resolves the
first problem, but the second problem persists even in this case.

The point is that if we consider the effects of
curved space, associated with the spherical surface at the event horizon, the
vortex core behaves very differently from the flat space vortex. 
Inside a flat space vortex, the symmetry is unbroken.  In this case therefore
one might imagine that the correct approximation is simply the unbroken theory,
i.e. the Reissner-Nordstrom solution.  However, the core of a long
straight vortex in flat space looks quite different from vacuum
electromagnetism in three
dimensions.  The only vacuum solution with the correct boundary
condition, associated with the turn-on of the Higgs field at the vortex surface
even if that surface is removed to infinity, involves a uniform magnetic 
flux in
the core (i.e. the vacuum solution in two dimensions).  In such a configuration
the gauge potential and magnetic field both behave quite differently from that
described in section 3. However, when the curvature associated with the
spherical event horizon is taken into account, as it must be when the string
core size is larger 
compared to the event horizon size (the thick string limit),
the presence of extra
$r^2$ contributions in the spherical derivatives around the event horizon 
allow a
vortex solution in which both the gauge potential and magnetic field fall off
exactly as in the unbroken 
3d theory inside the core, so that the physics inside
becomes 
largely insensitive to the boundary conditions associated with the Higgs
field behavior at the vortex surface.  In other words, as $v
\rightarrow 0$ 
the thick string limit smoothly approaches the Reissner-Nordstrom
case--there are no singular effects due to boundary conditions at infinity in
the curved space solution.

The previous argument can be made more explicit by examining the action and
equations of motion of the vortex, in both the thin and thick string limits.  
Define the quantities
\bea
y = {r \over r_{H}}, \,\,\,\, 
\ep^{2} = 2 e^{2} r_{H}^{2}v^{2}, \,\,\,\,
\beta_{0} = { \lambda \over 2e^{2}}
\eea
The thick string limit is taken by letting
$\ep^2$ tend to zero.  Using these variables
 the action of the vortex becomes:
\bea
S_{v} & = & { 1\over 4 \pi} { 2 \pi^{2} \over e^{2}}
\int_{1}^{\infty }
y^{2} dy e^{-\phi(y)} \mu_{y} 
\eea
where 
\bea
\mu_{y} & = &  (32 \pi e^{2}r_{H}^{4}) \, \mu \\
 & = &
 a'^{2}(y) e^{2\phi(y)} +
 \ep^{2} (1-{ b(y) \over y })^{-1} e^{2\phi(y)} 
  a^{2}(y) f^{2}(y) 
\eea
\vspace{-1.3cm}
\be
 + \,\,
 4 \ep^{2} (1-{ b(y) \over y }) f'^{2}(y) 
+ \beta_{0} \ep^{4} (f^{2} -1 )^{2} 
\label{eq:mu}
\ee
In the above $b|_{y=1} = 1$.
The other components of the energy 
momentum tensors can be similarly witten in terms of variable $y$,  
\bea
\rho & = T^{t}_{t} = & { 1\over 4 \pi}  { 1 \over 8e^{2}r_{H}^{4}} 
[ a'^{2}(y) e^{2\phi(y)} - 
 \ep^{2} (1-{ b(y) \over y })^{-1} e^{2\phi(y)} 
  a^{2}(y) f^{2}(y) \\
& & + \,\,4 \ep^{2} (1-{ b(y) \over y }) f'^{2}(y) 
+ \beta_{0} \ep^{4} (f^{2} -1 )^{2} ] 
\eea

The action and the components of  energy-momentum
tensor above are quite similar 
in form to those in the 
flat space.  However, there are two 
notable differences: the 
integral measure now becomes $y^{2} dy $ instead of $ydy$  
and there is an additional 
dimensionless parameter $\ep$ which measures the relative size
of the horizon and the vortex.  

In the thin string limit, $\ep^2 \rightarrow \infty$, 
from equation (\ref{eq:mu}),  we can see that $f$ tends to  be $1$ 
everywhere except possibly around the origin $y=1$. 
To find the leading piece, we can ignore the back-reaction of the 
vortex to the background geometry (since we assume $v \ll M_{pl}
\ll M_{BH}$.) and set $y=1$ whenever there
is an explicit dependence in the action integral. The 
simplified 
action is:
\bea
S_{v} & = & { 1\over 4 \pi} { 2 \pi^{2} \over e^{2}}
\int_{1}^{\infty } \,dy \,
[a'^{2}(y)  +
 \ep^{2} {1 \over y-1}
a^{2}(y) f^{2}(y) \\
& + &
 4 \ep^{2} (y-1) f'^{2}(y) 
+ \beta_{0} \ep^{4} (f^{2} -1 )^{2}] 
\eea
The results of the previous section  can be recovered 
when  we change the integration variable  in the action to  $x$, defined
as $x^{2} = 4 \ep^{2} (y-1)$, in terms of which,
the dependence of the action on $\ep$ factors out, and the action becomes:
\be
S_{v} = 4 \pi r_{H}^{2} T_{NO}
\label{eq:ac}
\ee
where $T_{NO}$ is the standard Nielsen-Oleson action for the vortex in 
two dimensional flat space. 
Note that in equation (\ref{eq:ac}), the prefactor $r_{H}^{2}$ comes 
from the factored-out $\ep^{2}$ piece and the $v^{2}$ in $\ep^{2}$ has
been absorbed into $T_{NO}$.
Equation (\ref{eq:ac}) is 
nothing but
\bea
S_{v} = ( {\rm Worldsheet \,\, Area}) T_{string}, 
\eea
as expected \cite{CPWa}.

As $\ep^2$ becomes smaller and smaller, the vortex becomes
thicker and thicker compared to the size of the  event horizon, and 
from equation (\ref{eq:mu}), it   
is clear  that as the $\ep^2$ goes to zero,  
the action reduces to that of 
Euclidean Reissner-Nordstrom black hole. To see that the
transition from the thick string limit to Reissner-Nordstrom
case is smooth, i.e. there are no singular effects encountered due to
boundary conditions from symmetry breaking at infinity, let us look at the 
equations of motion for the vortex. 
These
are: (again here we ignore the back-reaction of the vortex) 
\be
a'' + {2 \over y} a' = \ep^{2} (1-{1 \over y})^{-1}a f^{2} 
\label{eq:em1}
\ee
\be
{1 \over y^{2}}[y^{2} (1-{1 \over y}) f']' = 
{1 \over 4} (1-{1 \over y})^{-1} a^{2} f 
+ {1 \over 2} \beta_{0} \ep^{2} (f^{2}-1)f
\label{eq:em2}
\ee
The leading piece  of $a(y)$ can be found by 
solving equation (\ref{eq:em1}) with $\ep=0$, which is
$a(y)= 1/y$,  as expected \cite{CPWa}. Note the $y^{2}$ in the integral measure
has made the gauge potential $a(y)$ behave like a three dimensional
vacumm electromagnetic potential, resulting  an electric field falling
off as $1/r^{2}$. 
It is instructive here to recall the equation of motion for the vortex
in two dimensional flat space.
The equation for the gauge potential is:
\bea
a'' - {1 \over r} a' = f^{2}a
\eea
As we take $f=0$, the equation above reduces to that for electromagnetism
in two dimensional space, where $a \propto r^{2}$ and  the magnetic 
field (in this case given by $a'/r$)
is constant over space. This is the familiar vortex core behavior in flat
space.  If it persisted in curved space, then the thick string limit
would have a singularity associated with the surface at infinity which could
not be ignored.

For a small but non-vanishing 
$\ep$, we can estimate the range in which $a(y) \sim 1/y$ by looking
at  when the  term proportioanl to $\ep^{2}$ in  equation (\ref{eq:em1}) 
becomes  comparable to the other terms.
$$
a'' \sim \ep^{2} {y \over y-1} a f^{2} \,\, \Longrightarrow
\,\, 
{1 \over y^{3}} \sim \ep^{2} {y \over y-1} {1 \over y}
\,\, \Longrightarrow \,\, y \sim {1 \over \ep}
$$
where we have used the fact that $a \sim 1/y$ and 
$f \sim O(1)$. Thus the piece proportional to $\ep^{2}$
becomes important only  after $y$ is of the same order as or larger than
$1/\ep$, so that the range in which  $a$ behaves as  $1/y$ increases 
as $1/\ep$ with $\ep$ going  to zero, and the limit is indeed continous.
Note that while the term of the order $\ep^{2}$ 
in equation (\ref{eq:em1}) is proportional
$1/(y-1)$, an analysis of equation
(\ref{eq:em2}) implies that around $y=1$, $f \propto \sqrt{y-1}$, so around
$y=1$ the $f^{2}$ piece will cancel the divergence in $1/(y-1)$
so there is no singularity.

It might seem from this argument that one should ignore the details of
the $v$-dependent corrections to the black hole temperature in 
the thick string limit, as indeed was advocated by CPW.  However we have argued
in our previous work, and we demonstrate explicitly 
below, that this is not always
consistent.  Moreover, we will also demonstrate that while first order
corrections in $\ep^2$ to the temperature can dominate, the first
order contribution to the action is always negligible, and so the
instanton action is always greater than the Schwarzchild action, as is required
for consistency.

As the vortex becomes 
thicker and 
thicker with $\ep^{2}$ smaller
and smaller, the mathemtical content of the above discussion is that
in this limit there are no small or large parameters
other than $\ep^{2}$ in the action
and the energy momentum tensor.
We can thus expand these quantities in terms of $\ep^{2}$:

\bea
S_{matter} & = & S_{0} + \ep^{2} S_{1} + \cdots \\
\rho & = & \rho_{0} + \ep^{2} \rho_{1} + \cdots \\
\tau & = & \tau_{0} + \ep^{2} \tau_{1} + \cdots \\
\mu & = & \mu_{0} + \ep^{2} \mu_{1} + \cdots 
\eea

The contributions from the 
zeroth order terms are identical to those in the 
case of the
Euclidean Reissner-Nordstrom black hole, simply 
replacing $\omega$ by $2\pi$, so
that in this case the
Hawking temperature and the action are given by,
\bea
\bh & \approx & 8 \pi G M_{BH} 
[ 1  + ({1 \over 4 e})^{4} {1 \over G^{2}M_{BH}^{4}} 
+ \cdots ] \\
& \sim & 8 \pi G M_{BH} 
( 1 + O({M_{pl}^{4} \over M_{BH}^{4}}) + \cdots )
\eea
\bea
S & = & {(\bh)^{2} \over 16 \pi G} 
[ 1 + G ({ 2 \pi \over \bh e})^{2} ]^{2} \\
& = &  {(\bh)^{2} \over 16 \pi G} + {\pi \over 2 e^{2}}
 + { \pi G \over 4 e^{2}} ({ 2 \pi \over \bh e})^{2} \\
& \sim & {(\bh)^{2} \over 16 \pi G} + O(1) 
+ O[{M_{pl}^{2} \over M_{BH}^{2}}] + \cdots 
\eea
 
Here it is crucial that the lowest order correction to the  
Hawking temperature is of order 
$M_{pl}^{4}/M_{BH}^{4}$, instead of $M_{pl}^{2}/M_{BH}^{2}$ precisely
because the lowest order contribution in $G\Phi$ to the temperature
vanished in
the unbroken theory (c.f. eq.
(\ref{eq:emzero})).  Note that by comparison, the lowest order correction
to the action is of $O(1)$.

The contributions from the first order terms (in $\ep^2$) can be found
using eq (\ref{eq:HTf}) and (\ref{eq:app}).
\bea
2(A+B) - \frac{m}{M_{BH}} & = & \frac{ 8\pi G}{r_{H}} \ep^{2}
 \int_{r_{H}}^{\infty} dr
[ (4\mu_{1} r r_{H} - 2\mu_{1} r^{2}) 
- (\rho_{1} - \tau_{1})r(r-r_{H})] \\
& = & { G \over 4 e^{2} r_{H}^{2}} \ep^{2} T  
\eea
where 
\bea 
T = \int_{1}^{\infty} dy
[ (4\mu_{y1} y  - 2\mu_{y1} y^{2}) 
- (\rho_{y1} - \tau_{y1})y(y-1)]
\eea
and $ \mu_{y1},\rho_{y1},\tau_{y1}$ are dimensionless quantities
obtained from the
corresponding unsubscripted quantities by extracting an overall
dimensionful scaling factor
$1/(4\pi[8e^2r_{H}^4])$.

In this case the expressions for the Hawking Temperature
and the action are:
\bea
\bh & = & 8 \pi G M_{BH} 
( 1 - { G \over 4 e^{2} r_{H}^{2}} \ep^{2} T  + \cdots) 
\eea
\bea
S & = & \int d^{4}x \sqrt{g}\, ({\cal L}_{E} - {1\over 16\pi G}R)
+ {1 \over 2}(\bh)M_{BH} \\
& \approx  & {1 \over 2} \int d^{4}x \sqrt{g}\, (\rho + \tau) + 
{1 \over 2}(\bh) {\bh \over 8 \pi G} 
( 1 +  { G \over 4 e^{2} r_{H}^{2}} \ep^{2} T  ) \\
& \approx  & 
2 \pi \bh \int_{r_{H}}^{\infty} (\rho + \tau) r^{2}dr + 
{(\bh)^{2} \over 16 \pi G} +  { \pi \over 4 e^{2} }\ep^{2} T \\
& \approx & {(\bh)^{2} \over 16 \pi G} + { \pi \over 2 e^{2}}
+ { \pi \over 4 e^{2}} \ep^{2} T' \\
& \sim & {(\bh)^{2} \over 16 \pi G} + { \pi \over 2 e^{2}}
+ O({v^{2}M_{BH}^{2} \over M_{pl}^{4}}) + \cdots
\eea
where in the above $ \pi / 2 e^{2}$ comes from 
$\rho_{0}$ and $\tau_{0}$ and 
\bea
T' & = &\int_{1}^{\infty} y^{2}dy (\rho_{y1} + \tau_{y1}) + T \\
& = & \int_{1}^{\infty} dy [ 2(\tau_{y1} - \mu_{y1}) y^{2} + 
(4 \mu_{y1} + \rho_{y1} - \tau_{y1})y] \\
& \sim & 1
\eea

Now we can compare in detail the results from
the zeroth order and the first order terms to the Hawking
Temperature \cite{KLHprl}, and to the action.
The ratio of the first order in $\ep^2$ contribution 
 to the zeroth order in $\ep^2$ contribution to the Hawking Temperature  
(and also the ratio of the subdominant $\ep^2 \ne 0$ correction to
the subdominant $\ep^2 = 0$ corrections to the action) is  
$$
\gamma = {\ep^{2}M_{BH}^{2} \over M_{pl}^{2}} =
 v^{2}M_{BH}^{4}/M_{pl}^{6}
$$ 

Now, recall that the thick string limit is
$$
\ep^{2} = v^{2}M_{BH}^{2}/M_{pl}^{4} \ll 1.
$$
Recall that for the semiclassical analysis of black
hole thermodynamics to be meaningful, $M_{pl}^{2}/M_{BH}^{2}$ 
has to be very small.  For sufficiently massive black
holes, it is certainly possible for both $\ep^2 \ll 1$ and $\gamma \gg 1$ (i.e.
if $M_{pl}^{2}/M_{BH}^{2} \ll \ep^2$),
in which case the contributions
from the the first order terms in $\ep^2$ cannot be neglected. 
For example, if we keep $\ep^2$ fixed but
let $(M_{BH}/M_{pl})^{2} \rightarrow \infty $ (which requires also
making $v^2 \rightarrow 0$ ), then $\gamma \rightarrow \infty$, so that while
both the zeroth and first order contributions in $\ep^2$ to the 
Hawking Temperature go to zero, the
first order piece becomes arbitrarily large compared to the second. Stated
another way, the limit in which only the Reissner-Nordstrom 
piece is considered, as was done by CPW, is { \it not}
 the generic thick string
limit, but is rather the limit $\gamma \ll 1$. 

Note that in the same limit, the first order in $\ep^2$ piece to the
action dominates over the first order in $M_{pl}^{2}/M_{BH}^{2}$, both
of these terms are subdominant compared to the positive zeroth order piece 
$\pi/(2e^2) $ associated with the Reissner-Nordstrom action.  As a result, the
instanton action in the thick string limit is (independent of the magnitude
of $\ep^2$) always greater than the Schwarzchild action, as required for
the instanton semiclassical approximation to be stable.

It may seem somewhat surprising that for sufficiently large black holes
the first order terms in $\ep^2$ may become comparable or larger than the
zeroth order terms.  However, 
this is perhaps understandable
when one considers the dependence of the detailed vortex solution on the
curvature at the event horizon.  Recall that for a flat space vortex, the
field behavior inside the vortex core differs dramatically compared to that
outside the event horizon of the Reissner-Nordstrom solution.  It is only the
curvature at the spherical event horizon which allows the vortex 
instanton to approach the Reissner-Nordstrom  solution in the thick string
limit.  Now, for larger black holes the curvature at the event horizon becomes
progressively smaller.  While it may be true that the
$v
\rightarrow 0$ limit always goes smoothly to the Reissner-Nordstrom case,
independent of the black hole mass,
increasing the black hole mass reduces the curvature effects at the horizon
which are responsible for the domination of the Reissner-Nordstrom contribution
compared to the symmetry breaking contribution proportional to the vev of the
Higgs field. Thus, one might expect that the value of 
$\ep^2$ must be correspondingly reduced as the black hole mass increases
in order for the first order contribution in $\ep^2$ to be negligible. Put 
another way, the string must be correspondingly thicker (in relation to
the size of the event horizon) as the black hole
mass increases in order for the dominant
contribution to the Hawking temperature of the instanton to be that 
approximated
by the Reissner-Nordstrom solution.

Lastly we consider the possible sign of the first order term in the
expression for the black hole temperature given above.  It is clear
from the expression for $T$ that this is in general indeterminate.  However,
if we make the ansatz that the first order term takes a form similar
to that which would occur for the flat space vortex, which does not seem
unreasonable, then, assuming the first order quantities all have support
over a scale $r_s \gg r_{H}$  $ (y \gg 1) $

\bea
T  \rightarrow -  \int_{1}^{r_{s}} 
 (2\mu_{y1} + (\rho_{y1} - \tau_{y1})) y^2 \ dy\\
\rightarrow - 2/3 (r_s / r_H)^3 < \mu_{y1} >
\eea

This is manifestly negative, and hence the contribution of this piece
to the Hawking temperature would be of the same sign as the zeroth order
contribution.  To the extent that our anzatz is valid then, instantons in the
thick string limit would generically cool down a black hole---
the opposite of the
thin string result.  This result (obtained by CPW for purely the zeroth
order piece) is again heuristically understandable.  In the thick string limit,
because the instanton approaches the Reissner-Nordstrom solution, it
is to be expected that the effect on temperature will be that appropriate
to the Reissner-Nordstrom solution---namely to cool down the black hole.  Of
course, again because the actual dependence on discrete charge comes from
a sum over instantons and anti-instantons, the  effect of discrete
charge in this limit is to raise the black hole temperature.  This result,
which was a priori somewhat surprising, given the similarity of the thick
string limit and the Reissner-Nordstrom case, is then understandable as
being due to the similarity of the instanton and the Reissner-Nordstrom 
solution, combined with the fact that the sum over instantons
in the case of discrete charge produces results
in the opposite temperature correction compared to the single instanton
contribution.

\section{Non-Classical Electric Fields Redux} 

Our analysis thus far has enhanced our understanding
of the physical basis of the effects of instantons on black hole thermodynamics
by concentrating on the energy-momentum tensor outside the event horizon
associated with the instantons themselves.  Of course, as we have indicated
already, the limitation of this is first that it is not a single instanton but
rather
the sum over instantons which is finally relevant to the calculation of
physically measurable effects, and second, that instantons only give relevant 
corrections to quantities which otherwise have no perturbative contributions,
and which thus vanish in the classical limit. 

Nevertheless, while focussing on instantons, it is worth recalling that
the instanton solutions described here are not present in the Abelian-Higgs
theory in flat space.  Indeed, it is well known that 
there is no instanton in the
Abelian-Higgs model in four dimensional
flat spacetime since the Euclidean
sector  has the topology of $R^{4}$ or 
$R^{3} \times S^{1}$, which does not admit any finite energy solutions.
In a (1+1) dimensional spacetime, however, there are instantons which 
correspond to localized vortices in the $r- \tau$ plane.
Now, as we have discussed, the black hole sector of the 4-d
Einstein-Abelian-Higgs model has  the topology of
$R^{2} \times S^{2}$.  Thus, the system admits a finite action instanton solution
which corresponds to a vortex sitting in the 2-d $r-t$ plane ($R^{2}$) of a
black hole  with the other two dimentions $\theta , \phi$ ($S^{2}$) suppressed. 
(It  can  be imagined  as 
a Eulidean string with its worldsheet  a sphere $S^{2}$.      
Because the string worldsheet $S^{2}$ is compact, 
the solution has finite action.)  We have
seen that the existence of these new instantons  
can affect the properties of the black hole.  Moreover, besides
the effect on black hole thermodynamics, there is an even
more striking effect:  
the instanton sum results in a quantum mechanical electric field outside the 
event horizon of the black hole endowed with discrete hair \cite{CPWa,CPWb}.
 
In this section we wish to comment on this aspect of the Euclidean instanton
solutions described here, and their relation to instantons in the
two dimensional Abelian Higgs model.

As emphasized by CPW, the partition function for a  black hole 
in a charge $Q$
sector is given by\cite{CPWa}:
\be
Z(\beta, Q) = Tr (P_{Q} e^{-\beta  H}) =
\int_{-\infty}^{\infty} d\omega \,\, e^{-i2 \pi w Q/\h e}
\,Z(\beta, \omega)
\label{eq:par1}
\ee
where $P_{Q}$ is the charge projection operator and 
\[
Z(\beta,\omega) = \int_{\bh, \omega} dA\, d\phi \,\, 
\exp({-S_{E} \over \h})
\]
is the Euclidean path integral over configurations 
which are periodic in $\tau$ and satisfy the gauge constraint described
earlier:
\be
{e \over 2 \pi} \int_{0}^{\bh} dt \,\,A_{t} \,\,\,|_{r=\infty} = \omega
\label{eq:w1}
\ee
(Recall that in the Higgs phase $\omega$
is quantized, with only integer values allowed.  Thus the integral in
eq (\ref{eq:par1}) reduces to a summation over all integers.
In this case, we can see from the phase factor in   (\ref{eq:par1})
that only the fractional part of $Q/\h e$ is physically meaningful.) 
  
Restricting to the Euclidean $r-t$ plane, 
we can also write eq (\ref{eq:w1}) as,
\be
\omega = {e \over 2 \pi} \oint_{r-t}   \,\, \vec{A} \cdot \vec{dl}
=  {e \over 4 \pi} \int_{r-t} d^{2}x \ep_{\mu \nu} F_{\mn}
\label{eq:w2}
\ee
\[
\m, \n = 0,1
\]
where in the above $\vec{A}$ denotes  the $t,r$ components 
$(A_{0},A_{1})$ of 
the vector potential $A$. 
Plugging  (\ref{eq:w2}) into  (\ref{eq:par1}), 
we get,
\be
Z(\beta, Q) =   \int_{\bh} dA \, d\phi \,\, 
\exp({ -S_{E} \over \h} - 
i { e \theta \over 4 \pi}  \int_{r-t} d^{2}x \ep_{\mu \nu} F_{\mn})
\label{eq:par2} 
\ee
\[ 
\m, \n =0,1
\]
We have defined $\theta$ as
\be
\theta = {2 \pi Q \over \h e}
\label{eq:th}
\ee
and the integral is over 
all configurations periodic in $\tau$.   
(From the comment below eq (\ref{eq:w1}), it is clear
that in the broken phase, the theory is periodic in $\theta$
with period $2 \pi$.)
Eq (\ref{eq:par2}) shows that 
to get the partition function for a black hole in 
a charge $Q$ sector, instead of using the charge projection 
and $Z(\beta,w)$ in (\ref{eq:par1}), we can simply 
integrate over all configurations, provided we use a modified
$\theta$ dependent  action given by (up to gauge fixing),
\bea
S_{\theta} = { S_{E} \over \h} + i { e \theta \over 4 \pi}  
\int_{r-t} d^{2}x \ep_{\mu \nu} F_{\mn}, \,\,\,\,\,\,
\m, \n =0,1
\label{eq:to}
\eea
The additional term, 
is nothing but the familiar topological term in two dimensions with
$\theta$ as the topological charge.
Since the system is   spherically symmetric, we 
expect the  integration over  spherically symmetric 
configurations would dominate the integral (\ref{eq:par2}), 
in which case, the system then reduces completely to the 
corresponding model in (1+1) dimensions with a 
topological $\theta$ term.

It is interesting to see here that the electric charge $Q$ of 
a black hole
reduces, in leading
approximation, to a topological charge in (1+1) dimensions. 
While this correspondence seems surprising at first sight, 
it can be understood as follows.
The Coulomb phase of  a (4d) $U(1)$ theory   respects a 
superselection rule labeled by electric charge  $Q$ associated
with a long-range electric field. ($Q$ can be
discrete (integer multiple of some basic unit $e_{0} \h$) 
or continous 
depending on whether the  $U(1)$ symmetry is compact or not.)
In 
the Higgs phase, 
although classically the superselection rule no longer 
holds  due to the screening of the long-range electric field, 
quantum  mechanically there is still a nontrivial 
superselection rule left, provided that the original $U(1)$ is 
non-compact, or,   in the compact case,  
the condensation charge $e = N e_{0}$ with $N$ an integer
other than $1$.
The  superselection sector now is 
labeled by $Q (mod \,\h e)$, associated with a nontrivial
Aharonov-Bohm phase 
$$ \exp( {2 \pi i  Q \over \h e})$$ 
at infinity. In terms of $\theta$ defined in (\ref{eq:th}),
in the noncompact case, the system is labeled by a continuous
parameter 
$$\theta = {2 \pi Q \over \h e} \in (0,2 \pi),$$ 
while in the compact
case, it is a $Z_{N}$ charge, 
$$\theta \in Z_{N} = \{ \exp(2 \pi k/N), \,\, k=0,1,\cdots N-1\}.$$
Now the two-dimensional   system (with an  unbroken or broken $U(1)$) also 
respects a superselection rule where superselection sectors 
are labeled by the topological charge $\theta$. The different
superselection sectors fall into  different $\theta$-vaccua which 
are accompanied by a constant  
background electric field (classical in the unbroken
$U(1)$ theory, nonclassical in the broken $U(1)$).
Now the correspondence between the electric charge
$Q$ and the topological charge is clear: from a 4-d 
point of view, the superselection sector the black hole 
falls into is labeled by its electric charge $Q$. 
On the other hand, since the system is spherically 
symmetric, the theory reduces to a effective 
2d theory, where one can label the
superselection sector by the topological charge associated
with the topological term (\ref{eq:to}).

The  correspondence between the black hole system here
and 2-d models can also help us gain some insights
into the existence of a quantum
mechanical electric field outside the event horizon of 
a black hole with quantum hair first discovered by
CPW\cite{CPWa,CPWb}.
It just corresponds to the familiar fact that 
there is a  nonclassical background 
electric field in the (1+1) dimensional abelian-higgs 
model associated with instantons and the topological term.
There is a notable 
difference, however. In the (1+1) dimensional Abelian Higgs model, 
there is a dilute instanton gas---the instantons 
can sit at any point in the two dimensional Euclidean space---
resulting in a constant electric field as the system (due to  
translational invariance).
In the quantum hair case, however,
since the existence of the instanton solutions 
depends crucially on the  topology of a black hole, they
can only sit at the event horizon of a black hole.
Thus, instead of a dilute instanton gas, 
we only have a single localized 
instanton, resulting  a localized electric field 
which  dies off at large distance from the
event horizon.

This exact analogy between discrete quantum numbers on black holes
in four dimensions and topological quantum numbers on related two dimensional
systems may be of some interest for those wishing to interpret black hole
entropy in terms of underlying quantum numbers associated with state counting
in theories in which the black hole is the low energy limit of a string
theory.
 
\section{Conclusions}

Our analysis, based on considerations of the energy momentum tensor of 
fields outside the event horizon has allowed us to calculate 
instanton corrections to the temperature of black holes.    
Utilizing this formalism, the
competition between different terms in the energy momentum tensor is
explicitly demonstrated to lead to either a heating up, or a cooling down of 
the
black hole in a way which is physically transparent.  In addition, we see 
how the
Euclidean energy momentum tensor associated with the instanton fields outside
the event horizon yields semiclassical corrections which can violate the weak
energy condition. In the thin string limit our results are identical
to those obtained by CPW.  However, our analysis
has allowed us to extend their results. In the
thick string limit, the effects of symmetry breaking need not be sub-dominant,
allowing us to reconcile our flat space intuition with the curved space
results.  This new correction probably does not result in a
qualitative change in the picture, since our intuition also suggests that the
first order contributions will have the same sign as the zeroth order
contributions, but this remains to be quantitatively checked.

We stress again that the corrections which we calculate are themselves only
relevant to quantities which vanish in the classical limit.  In the case
of discrete quantum hair, the instanton contributions are relevant because
(a) there is no perturbative or classical signature associated with the
quantum hair, and (b) the quantum phases associated with their contributions to
Euclidean path integrals allow a non-zero result when a sum over instantons
and anti-instantons is performed, as demonstrated by CPW.   Nevertheless, our
results indicate that one may fruitfully extend Minkowski space methods
designed to probe the effects of classical fields outside the event horizon on
the thermodynamical properties of black holes to the Euclidean regime of
semiclassical phenomena.  This allows a more
intuitive physical picture of the origin of such effects.  It may also be
useful in exploring the nature of other semiclassical contributions to black
hole thermodynamics beyond those considered here associated with quantum hair. 

It would be very nice to be able to extend our analysis so that the Euclidean
field outside the event horizon associated with the instanton sum could be
treated directly, so that perhaps the relation between the non-classical
electric field outside the event horizon and the change in the black hole
temperature could be directly linked. 

Finally, returning to the traditional Euclidean partition function in
the instanton approximation, we have been able to derive an exact analogy
between discrete charge on black holes in four dimensions and topological
charge in related two dimensional systems.  Using this analogy, the
non-classical electric field outside black holes endowed with discrete hair
can be understood as merely a special case of the well known existence of a
non-classical electric field in the spontaneously broken two dimensional
Abelian Higgs model endowed with a topological term.  This analogy may be
useful in considerations of the relation between black hole entropy and state
counting.

\bigskip
LMK would like to thank Frank Wilczek for conversations early on in
this analysis, and Sidney
Coleman and John Preskill for several long and very illuminating
conversations which helped steer us away from several red herrings.
Finally, LMK and HL thank Tanmay Vachaspati for discussions of strings in
curved space and key insights which helped us resolve the details of the thick
string limit. LMK and HL are supported by the DOE and funds from Case Western
Reserve University.

\end{document}